\documentclass{mn2e}
\usepackage{epsfig}
\usepackage{epsf}
\usepackage{graphicx}

\title[Periastron Collisions]{Explosions Triggered by Violent
  Binary-Star Collisions: Application to Eta Carinae and other
  Eruptive Transients}

\author[Smith]{Nathan Smith\thanks{Email:
    nathans@as.arizona.edu} \\ Steward Observatory, University of
  Arizona, 933 North Cherry Avenue, Tucson, AZ 85721, USA}

\begin{document}
\date{Accepted 0000, Received 0000, in original form 0000}
\pagerange{\pageref{firstpage}--\pageref{lastpage}} \pubyear{2002}
\def\arcdeg{\degr}
\maketitle
\label{firstpage}

\begin{abstract}

  This paper discusses a model where a violent periastron collision of
  stars in an eccentric binary system induces an eruption or explosion
  seen as a brief transient source, attributed to luminous blue
  variables (LBVs), supernova (SN) impostors, or other transients. The
  key ingredient is that an evolved primary increases its photospheric
  radius on relatively short (year to decade) timescales, to a point
  where the radius is comparable to {\it or larger than} the
  periastron separation in an eccentric binary.  In such a
  configuration, a violent and sudden collision would ensue, possibly
  leading to substantial mass ejection instead of a binary merger.
  Repeated periastral grazings in an eccentric system could quickly
  escalate to a catastrophic encounter.  Outbursts triggered by tidal
  disturbances or powered by secondary accretion of the primary star's
  wind have been suggested previously.  Instead, this paper proposes a
  much more violent encounter where the companion star plunges deep
  inside the photosphere of a bloated primary during periastron, as a
  result of the primary star increasing its own radius.  This is
  motivated by the case of $\eta$~Carinae, where such a collision must
  have occured if conventional estimates of the present-day orbit are
  correct, and where peaks in the light curve coincide with times of
  periastron.  Stellar collisions may explain {\it brief} recurring
  LBV outbursts like SN~2000ch and SN~2009ip, and perhaps outbursts
  from intermediate-mass progenitor stars (i.e., collisons are not
  necessarily the exclusive domain of very luminous stars), but they
  cannot explain all non-SN transients.  Finally, mass ejections
  induced repeatedly at periastron cause orbital evolution; this may
  explain the origin of eccentric Wolf-Rayet binaries such as WR~140.

\end{abstract}


\section{INTRODUCTION}

Considerable mystery surrounds the class of transients that includes
giant eruptions of luminous blue variables (LBVs) and other so-called
supernova (SN) impostors.  These are thought to be non-terminal
eruptions of massive stars, although recent evidence suggests that
similar eruptions occur in evolved intermediate-mass stars (initial
masses $\la$8 $M_{\odot}$) as well.  These eruptions have a diverse
range of peak absolute magnitude, fading rate, total energy, spectral
morphology, and progenitor initial masses (see Smith et al.\ 2010b for
a recent discussion of members of the class).  So far, there is no
plausible theory to explain these outbursts.

Included among these non-SN outbursts are very brief events, which
reach peak absolute magnitudes of $-$12 to $-$14, but which only last
a few days or weeks, in contrast to other LBV eruptions that can go on
for years.  Some of these brief events seem to repeat: Multiple
$\sim$100 day events were seen before the eruption of $\eta$ Car
(Smith \& Frew 2010), numerous rapid spikes were seen before the
eruption of SN~1954J (Tammann \& Sandage 1968), and in modern times
both SN~2000ch and SN~2009ip have shown recurring rapid brightening
and fading (Wagner et al.\ 2004; Pastorello et al.\ 2010; Smith et
al.\ 2010a; Drake et al.\ 2010).  The repetition of events is
obviously suggestive of binary encounters, but this is quite
speculative for extragalactic eruptions where we have limited
information about the progenitor systems.


However, $\eta$~Carinae is a unique nearby case, known to have
survived to the present day in a binary system with reasonable
estimates of the orbital parameters. We know the approximate amount of
mass ejected in the eruption by measuring its circumstellar nebular
mass, and fortunately, we have a good historical record of the
observed brightness during the event as well.

Recently, Smith \& Frew (2010) presented over 50 newly recovered
historical estimates of the visual magnitude near the peak of the
mid-19th century eruption.  They demonstrated clearly that the Great
Eruption was not a simple 15-20 yr brightening of the star as is often
assumed.  Instead, there were multiple brief 100-day peaks leading up
to the eruption.  Smith \& Frew (2010) showed that these brief peaks
occurred within weeks of periastron, providing that the orbital period
was slightly smaller at that time than the period measured today; this
should be the case, since the system ejected 15 $M_{\odot}$ during the
event (Smith et al.\ 2003).

It is hard to believe that major brightenings occuring repeatedly so
close to times of periastron would be a coincidence.  This begs the
question: ``What actually happened during the periastron encounters of
Eta Carinae in the early 1800s?''  In \S 2 we consider the parameters
of $\eta$~Carinae, and show that a stellar collision must have occured
at periastron before and during the Great Eruption, where the
presumably main-sequence O-type secondary star plunged deep inside the
effective photosphere of the bloated primary.  This is certainly a
violent and exotic encounter, spurring several other questions that we
attempt to address briefly: What are the physical and observable
effects of one star plunging into another's envelope and emerging?
What is the energy budget in such an event?  What causes the primary
star's radius to increase so quickly, and what other types of systems
might experience this?  If the system failed to merge and has survived
as a binary today, then how does such an encounter affect its orbital
evolution?  We hope that ideas outlined here will help guide numerical
models of such an encounter.  In a subsequent paper, we will discuss
possible implications for the structure of $\eta$ Car's nebula.

\begin{figure*}\begin{center}
\includegraphics[width=5.1in]{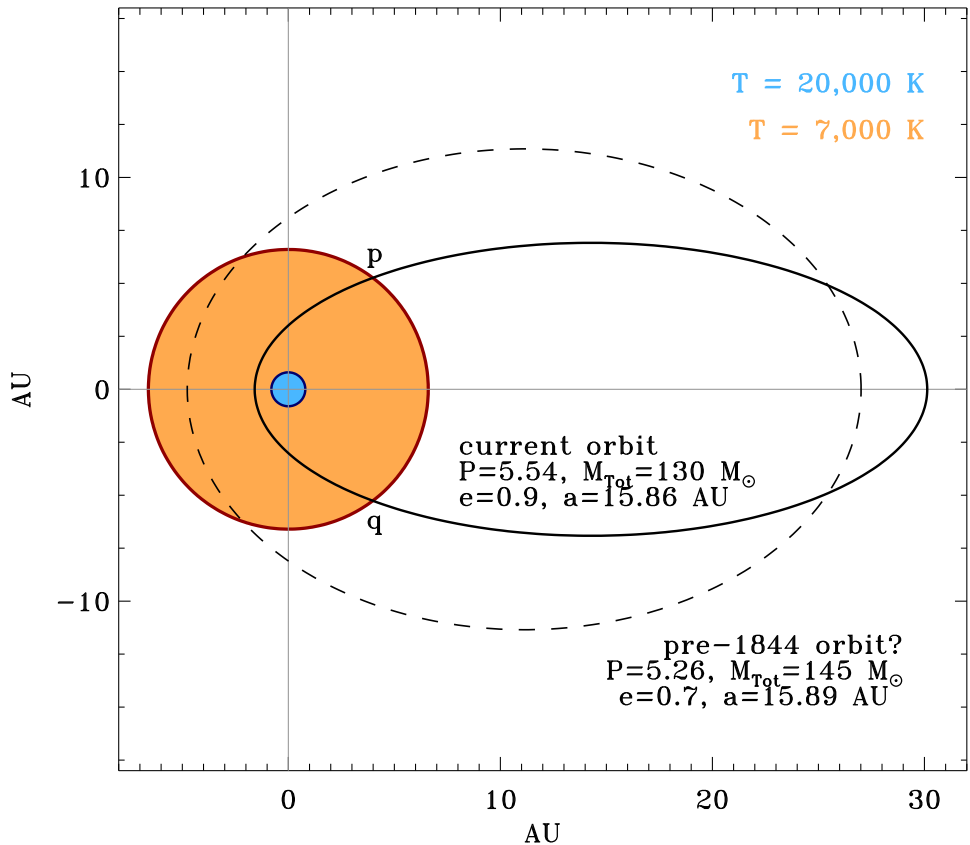}
\end{center}
\caption{A sketch of the $\eta$ Carinae binary system.  The solid
  ellipse uses conventional parameters for the present-day orbit
  derived from model fits to the X-ray light curve (i.e., $e$=0.9,
  although some models have proposed even higher eccentricity).  The
  dashed ellipse is a hypothetical orbit before the major mass
  ejection of the Great Eruption when $\sim$15 $M_{\odot}$ was lost
  from the system; we do not know the pre-1844 eccentricity, so we
  show a hypothetical lower value of $e$=0.7 for illustrative
  purposes.  The smaller blue circle shows the effective photospheric
  radius of the primary in its present-day (and presumably its
  pre-1600 A.D.) state with $T_{\rm eff} \simeq$ 20,000~K and
  L$\simeq$4$\times$10$^6$ $L_{\odot}$.  The larger orange circle is
  the primary star's radius for the same luminosity and $T_{\rm eff}
  \simeq$ 7,000~K.  Clearly, a violent collision would have occurred,
  placing the secondary well inside the primary star's photosphere.}
\label{fig:orbit}
\end{figure*}

\section{ETA CARINAE}

In the conventional picture of LBV outbursts of the normal
S~Doradus-type, a star will brighten at visual wavelengths by an
amount comparable to its bolometric correction (BC), as it transitions
from a hot quiescent state (usually very late O-type or early B-type)
to its cool state as an F-type supergiant with an apparent temperature
around 7000--8000~K.  While this does not appear to hold for all giant
LBV-like eruptions (Smith et al.\ 2010b), this paradigm seems well
established for conventional S Doradus-type excursions of LBVs
(Humphreys \& Davidson 1994).

By transitioning to cool temperatures at roughly the same luminosity,
an LBV must increase its emitting radius by a huge factor, $R \propto
T^{-2}$.  For a typical transition from early B to an F-type
supergiant, the radius may increase by a factor 7--10.  Whether this
is a true increase in the star's hydrostatic radius (Groh et al.\
2009) or the radius of a pseudo-photosphere in the opaque wind
(Davidson 1987) is controversial.  The important point, though, is
that any release of energy occurring inside the optically thick
(pseudo)photosphere will be thermalized.

This {\it large and sudden increase in radius} is the most crucial
physical change in the system, for the model discussed in this paper.
In an LBV that is a single star or a very wide binary, this S~Doradus
excursion will simply cause the star to brighten at visual
wavelengths, on typical timescales of years to decades as seen in the
standard Hubble-Sandage variables (Hubble \& Sandage 1953; Humphreys
\& Davidson 1994).  However, if the LBV happens to reside in a binary
system with a period of a few years and with some eccentricity, the
huge increase in radius can be cataclismic.

Consider what happens if we apply this entirely conventional picture
of S~Doradus variability to $\eta$~Carinae.  In the century before the
Great Eruption, $\eta$ Car steadily brightened from an apparent visual
magnitude of about 3.5 to 1.2 mag (Frew 2004; Smith \& Frew
2010).\footnote{Note that putative fluctuations between 2nd and 4th
  mag are from upper and lower limits.  Reliable reports of
  $\eta$~Car's magnitude are consistent with a steady brightening
  during the 18th century (see Smith \& Frew 2010).}  The simplest
interpretation of this $\sim$2.3 mag brightening is a long S~Dor-type
excursion where the star remained at constant luminosity, but
brightened at visual wavelengths by an amount equal to its BC.  A BC
of 2.3 mag corresponds to a temperature of around 20,000~K, or an
approximate spectral type of B1~Ia.  This is roughly consistent with
the star's present-day state (e.g., Hillier et al.\ 2001).  The
current bolometric luminosity of $\eta$~Car is about 4$\times$10$^6$
$L_{\odot}$ (Smith et al.\ 2003), allowing for a 10--20\% contribution
to the total luminosity from a companion star.  At constant $L$, its
photospheric radius must then have increased from $\sim$170
$R_{\odot}$ (0.8 AU) to about 1400 $R_{\odot}$ (6.6 AU).

In the present-day orbit, models for the X-ray colliding wind emission
and other data suggest the following orbital parameters: $e$=0.9,
$P$=5.54 yr, a=15.9 AU, $M_1$=100 $M_{\odot}$, and $M_2$=30
$M_{\odot}$ (Parkin et al.\ 2009; Okazaki et al.\ 2009; Pittard \&
Corcoran 2002; Corcoran 2005; Mehner et al.\ 2010).  The closest
periastron separation between the two stars in this orbit is only 1.6
AU. Figure~\ref{fig:orbit} illustrates the obvious problem here.  By
the time $\eta$ Car brightened to its observed early 19th century
magnitude, its characteristic emitting radius was substantially larger
than the periastron separation in the binary system we see today.  In
other words, {\it the secondary star would have plunged deep inside
  the photosphere of the primary star at periastron}.  We do not know
the pre-1844 eccentricity, but Figure~\ref{fig:orbit} shows that a
violent collision would still occur even for a hypothetical
eccentricity as low as $\sim$0.7 (dashed ellipse).  This is a rather
exotic state of affairs, with implications discussed below.

Smith \& Frew (2010) have demonstrated that the brief brightenings of
$\eta$~Car in 1838 and 1843 occurred within weeks of periastron, if
the pre-1844 orbit is $\sim$5\% shorter than that observed today due
to mass loss from the system.  The indication of
Figure~\ref{fig:orbit} is that this is no mere tidal interaction of
two close stars, but a brutal collision where one star burrowed deep
inside the other star's bloated envelope.  The duration of the
periastron collision itself is the time for the secondary to move from
point $p$ to $q$ in Figure~\ref{fig:orbit}, which in this case is a
few months.  This is, interestingly, comparable to the duration of the
X-ray outbursts seen in the present-day colliding-wind binary.  It is
also comparable to the $\sim$100 day duration of the brief brightening
events in 1838 and 1843 (Smith \& Frew 2010).  Based on this, we
hypothesize that the brief 1838 and 1843 brightening events, where
$\eta$~Car's bolometric luminosity increased for a short time, were
the direct observed result of this collision.  The energetics of such
an encounter are discussed next.

\section{THE ENERGY BUDGET}

What actually happens energetically when one star suddenly plunges
deep inside the other's envelope?  Real models for such an encounter
do not yet exist.  A conventional assumption in models is that when
stars are in a close binary, the orbits will become tidally locked and
circularize, leading to either mass-transfer or a merger in a thermal
timescale.  The case of $\eta$ Car is quite different because the
primary star's radius increased dramatically on a timescale of a few
orbits, and the orbit is very eccentric, so circularization would not
occur.  Clearly the system survived the encounter and did not merge
into a single star, because we see an eccentric binary today, and the
times of periastron coincide with brightening events during the Great
Eruption.  Instead, the high orbital velocity at such a close
periastron separation evidently permitted the seconday star to plow
through the primary's envelope and escape out the other side.

Let us now consider the energy budget during such a collision.  We
take the duration of the collision to be $\sim$100 days, as noted
above.  During the brief 1838 and 1843 events, $\eta$ Car brightened
by about 1.6 mag, and therefore radiated an extra
$\sim$5$\times$10$^{47}$ erg beyond what the star would have radiated
anyway at its quiescent luminosity.  If we attribute the ejection of
the Homunculus nebula to these periastron collisions, which is not
necessarily the case, then the energy budget climbs to $\sim$10$^{50}$
erg because of the kinetic energy involved (Smith et al.\ 2003).  We
can evaluate a few hypothetical sources of energy in such an
encounter:

{\it Kinetic heating of the envelope}.  When one star plunges into
another's envelope, friction will drain the kinetic orbital energy of
the intruder and this will heat the envelope.  The available energy is
necessarily some small fraction of the total orbital potential energy
(about 7$\times$10$^{48}$ erg in this case), since there were multiple
encounters and the system did not merge.  The radiated energy was
about 10\% of the available orbital energy, so frictional heating of
the primary star's envelope is at least a plausible explanation for
the increase in luminosity during the brief 1838 and 1843 events.  It
cannot, however, explain the radiated energy budget of the entire
eruption or the kinetic energy of the ejecta, so some other source
must have powered the radiation of the main eruption and launched the
Homunculus.

{\it Radiation emitted by the secondary star}.  The secondary star
itself is less luminous than the primary, so its own radiation trapped
inside the primary star's envelope offers no relevant contribution to
the energy budget.  However, there is another consideration, mentioned
next.

{\it Accretion of primary star's envelope onto companion star}. In a
series of papers, Soker (see Kashi \& Soker 2010, and references
therein) has advocated a rather complicated model where a secondary
star accretes material from the wind of the erupting primary at
periastron.  In this picture, the direct accretion luminosity of the
secondary powers the extra radiation (for the full 15 yr duration of
the Great Eruption), and bipolar jets associated with the accretion
disk form the Homunculus nebula.  The model is analytic and involves
many unverified assumptions and assertions, and adopts huge primary
mass-loss rates as a precondition.  It is neverthelss worth
considering, but the situation must be quite different from that
envisioned by Soker et al. in several key respects: (1) instead of
accretion from the primary star's wind, the secondary would be {\it
  inside the primary} and would accrete directly from the primary
star's envelope (this affects the calculated Bondi-Hoyle accretion
rate, since the envelope is probably static, and not an accelerated
wind.  (2) Any radiation of the accretion luminosity from the
secondary would occur inside the photosphere of the primary, and would
therefore be absorbed and reprocessed by it. Thus, the radiation we
observe cannot be direct radiation from the accreting companion. (3)
The observed periastron event was observed to last only $\sim$100
days, so the accreted mass, total radiated energy from accretion, and
time over which accretion operated must have been much smaller than
calculated by Soker, who assumes that it occurs over $\sim$10 yr and
powers the full radiation of the Great Eruption.  With much less mass
accreted, the effect on the orbital evolution is less severe, and the
mass accretion budget becomes more reasonable.  (4) Hypothetical jets
launched by the accretion disk around the secondary would need to
drill through the primary star's envelope, so it is not clear that the
jets would survive.  Instead, they might simply impart their energy to
the primary star's envelope and induce a sudden (bipolar?)  explosion.
This is of course very speculative, but the point is that the
situation is quite different from an undisturbed accretion disk around
a secondary star that blows collimated jets.

{\it Induced mixing of fresh fuel into deeper layers}.  Dessart et
al.\ (2009) has explored the possibility of explaining observed
properties of some LBV-like transients with the deep and sudden
deposition of energy that induces an explosion.  One hypothetical
source for this is the sudden nuclear combustion of only 0.01--0.1
$M_{\odot}$ of fresh fuel mixed down into a deeper burning layer in
the star.  Massive stars may become unstable enough to do this on
their own at convective boundaries (see Meakin \& Arnett 2007), but if
a 30 $M_{\odot}$ star plunges deep inside the envelope of a 100
$M_{\odot}$ star that is near the Eddington limit anyway, one might
wonder if the ensuing disturbance could also trigger the necessary
small amount of mixing. The density gradient at a convective core
boundary is a formidable obstacle, but exploring the consequences of
such an event may be interesting.

Although it may verge on overspeculation, this last mechanism has some
advantages over the accretion model.  Since the accretion onto the
secondary can only cause the ejection of material in the outer
envelope at a point where the binding energy is low, it is difficult
to see how it could lead to the ejection of more than 10 $M_{\odot}$
and 10$^{50}$ ergs, as required for the formation of the Homunculus.
Instead, deposition of energy at a depth corresponding to a binding
energy of 10$^{50}$ erg and where a larger mass reservoir is available
seems like a more natural explanation (Dessart et al.\ 2009), which
can potentially be achieved in an explosive burning scenario.
Although models for triggering such an event have not yet been
explored, these leading comments are perhaps justified, given the
violent and exotic nature of such an encounter, plus the
well-established observational basis that it did in fact occur in
$\eta$~Carinae.

\section{THE PRIMARY STAR'S RADIUS, AND APPLICATION TO OTHER
  TRANSIENTS}

Observationally, $\eta$~Car's photospheric radius clearly increased in
the two centuries leading up to 1844 (see Smith \& Frew 2010).  In
such a very luminous and unstable system, one can plausibly attribute
this to the inherent instability of a star flirting with the classical
Eddington limit, as conventionally discussed for LBVs.  In fact, the
pre-1840s secular brightening could be attributed to a slow but
otherwise normal S~Doradus excursion, causing a change in the BC as
discussed above.  One might therefore expect other LBVs to be readily
able to experience violent periastron collisions, if they happen to be
in eccentric binaries.  Observations of the solar neighborhood suggest
that $\sim$10\% or more of binary systems have high initial
eccentricity above $e$=0.4 (e.g., Mayor \& Mermilliod 1984), although
this distribution is not well known for massive stars.  (The initial
eccentricity distribution is needed to estimate the expected rates of
periastron collisions.)

In that case, we may have a potential explanation for the very brief
and repeated brightening events seen in SN~2009ip and SN~2000ch (Smith
et al.\ 2010a; Pastorello et al.\ 2010; Drake et al.\ 2010).  Both
systems showed a rapid brightening and fading on time scales of
several days -- much quicker than one normally attributes to eruptions
of LBVs.  These brief episodes could plausibly be explained as
periastral grazings or true collisions due to the primary star's
increasing radius during S~Dor excursions.  The LBV instability is
notoriously irregular, occurring on year to decade timescales.
Depending on the orbital separation, periastron collisions may only
occur when the star increases its radius to the maximum brightness in
an S~Dor event.  Therefore, the appearance of sudden brightenings in a
system which did not previously exhibit it --- or in fact, the
irregular disappearance and reappearance of brief eruptions -- can be
explained for LBVs in eccentric binaries.  In other words, one can
expect periodic repetition of periastron grazings and collisions, but
{\it only when the primary is in an outburst state with an expanded
  radius}.  The repeated eruptions may therefore not be strictly
periodic because in some cycles, nothing observable will happen at
periastron.  Even so, an orbital period of around 190 days would
provide satisfactory coincidences in the case of SN~2000ch, judging by
the light curve from Pastorello et al.\ (2010).  As discussed in the
next section, the first appearance of such behavior may be brought on
by a runaway instability, leading to catastrophic encounters and
subsequent mass ejections.  When enough mass is lost, the envelope
will contract again, thereby shutting off the repeated collisions.

Other stars besides LBVs will increase their photospheric radius
during their evolution, although the typical few-year timescale for
S~Doradus variations of LBVs is well-suited to a sudden change in
radius during an orbit.  Imagine, for example, a star with an initial
mass of $\sim$7 $M_{\odot}$, born in an eccentric binary with a period
of a few years.  As a star works its way up the final asymptotic giant
branch (AGB), one could imagine sudden encounters if times of
periastron coincided with a major pulsation, for example.  If these
happen rather suddenly in an eccentric system, it may lead to a
collision and mass ejections rather than mass transfer or merger.
Perhaps this is an explanation for the LBV-like tranients that seem to
arise from relatively low-mass progenitors, like M85-OT, SN~2010U,
V838~Mon, etc., which bear many similarities to the eruptions of known
LBVs (see extensive discussion in Smith et al.\ 2010b, and references
therein).  Collisions and eruptions might also occur in the years
immediately preceding a core-collapse SN, if the final burning phases
trigger a rather sudden increase in the progenitor's radius, the
implications of which are potentially important for understanding
SNe~IIn and Ibn.

The observed case of $\eta$~Car demonstrates that such a stellar
collision will not necessarily lead to the successful merger of the
pair of stars.  Stellar mergers have already been suggested as
potential explanations for objects like M85-OT and V838 Mon (e.g.,
Kulkarni et al.\ 2007; Tylenda 2005), but the non-merger collisions
proposed here might be another viable explanation.  Indeed a merger
may occur in some cases, but the collision may instead trigger severe
explosive mass ejection, which may leave the binary system less bound,
and may produce a brilliant transient source in the process.  With
different orbital periods, eccentricities, and stellar radii,
different masses of the bloated primary envelope, and different
companion masses (not to mention the possibility of {\it compact}
companion stars), one can quickly imagine a wide diversity of ejected
mass, energy, and luminosity for the resulting transients.  This may
provide an attractive explantion for the huge diversity in LBV-like
eruptions and related transients observed so far (see Smith et
al. 2010b), but real models of such an encounter are needed.

\section{ORBITAL EVOLUTION}

Lastly, we briefly mention one more consequence of the type of stellar
collision described above.  If these events induce significant
ejections of mass from the system --- and if the mass ejection is
concentrated at periastron --- then multiple such encounters could
drive rapid orbital evolution.  Steady mass loss (as in normal stellar
winds) will tend to circularize and widen an orbit over time, but mass
loss events concentrated at periastron in an eccentric system will
tend to make the system more eccentric.

One can see that initially grazing encounters could potentially
escalate quickly to catastrophic collisions.  In a mildly eccentric
system with a periastron separation not much larger than the primary
star's radius, tidal friction and deposition of energy into the
primary star's envelope may be small at first (e.g., Moreno et al.\
1997).  However, it may initiate a feedback loop where successive
encounters disturb and inflate the primary star's envelope, making
each subsequent encounter more severe until a true collision is
unavoidable.  This provides an attractive explanation for the building
instability in the few years before a giant LBV eruption, as seen in
$\eta$~Car, SN~1954J, SN~2009ip, UGC~2773-OT, and perhaps HD~5980 (see
Smith et al.\ 2010b).

If these sorts of periastron collisions induce enough mass loss from
the system to completely remove the massive primary star's H envelope
and thereby form a Wolf-Rayet (WR) star, it would profoundly change
the orbit.  In particular, each successive periastron mass ejection
would leave the system less bound and more eccentric.  This type of
scenario may therefore provide a reasonable explanation for the origin
of very eccentric WR+OB colliding-wind binary systems like WR~140,
which has $e$=0.88 (Marchenko et al.\ 2003).  If the periastron mass
loss is severe enough, in some cases it may even unbind the system
altogether, forming an apparently single WR star.  The $\eta$~Car
system will likely be left unbound if it encounters one more mass-loss
event as extreme as the 1840s Great Eruption, providing that this mass
loss occurs at periastron.




\end{document}